\documentclass[preprint,times,12pt]{elsarticle}

\usepackage[german,american]{babel}

\usepackage[T1]{fontenc}

\usepackage{amsmath} 
\usepackage{amssymb} 
\usepackage{amsxtra} 
\usepackage{bm} 
\usepackage{mathtools}

\usepackage[OMLmathsfit]{isomath}

\def\ssans#1{\textsf{\textsl{#1}}}
\usepackage{miller}

\usepackage[usenames]{color}
\definecolor{darkgreen}{rgb}{0.0,0.5,0.0}
\definecolor{BurntOrange}{rgb}{0.8,0.3,0.0}

\usepackage[dvips]{epsfig}
\usepackage[dvips]{graphicx}
\usepackage{psfrag}
\usepackage{subcaption}

\usepackage[load-configurations=version-1]{siunitx}

\usepackage{float}

\usepackage{caption}
\usepackage{epic,eepic}

\usepackage{listings}
\usepackage{soul}

\usepackage[colorlinks=true, allcolors=blue]{hyperref}

\usepackage{graphicx} 

\journal{}

\begin{document} 

\title{Combined modeling and experimental characterization of Mn segregation 
and spinodal decomposition along dislocation lines in Fe-Mn alloys}

\author{Jaber Rezaei Mianroodi$^{1}$, 
Pratheek Shanthraj$^{2}$, 
Alisson Kwiatkowski da Silva$^{1}$, 
Bob Svendsen$^{1, 3}$, 
Dierk Raabe$^{1}$} 
\address{$^1$Microstructure Physics and Alloy Design, 
Max-Planck-Institut f\"ur Eisenforschung, D\"usseldorf, 
Germany}
\address{$^2$The Department of Materials, 
The University of Manchester, Manchester, UK}
\address{$^3$Material Mechanics, RWTH Aachen University, Aachen, Germany}

\corref{jaber.rezaeimianroodi@rwth-aachen.de}

\begin{frontmatter}

\begin{abstract} 

In the current work, Mn enrichment at dislocations in Fe-Mn alloys due 
to segregation and spinodal decomposition along the dislocation line is 
studied via modeling and experimental characterization. 
To model these phenomena, both finite-deformation microscopic phase-field 
chemomechanics (MPFCM) and Monte Carlo molecular dynamics (MCMD) 
are employed. MPFCM calibration is carried out with the same Fe-Mn MEAM-based potential used in MCMD, as well as CALPHAD data. 
Simulation results for Mn segregation to, and spinodal decomposition along, 
straight screw and edge dislocations as well as dislocation loops, are 
compared with characterization results from atom probe tomography (APT) for two 
Fe-Mn alloy compositions. 
In contrast to classical Volterra (linear elastic) dislocation theory, 
both MPFCM and MCMD predict a non-zero hydrostatic stress field in screw cores. 
Being of much smaller magnitude than the hydrostatic stress in straight edge 
cores, much less solute segregates to screw than to edge cores. 
In addition, 
the segregated amount in screw cores is below the critical concentration of 0.157 for 
the onset of spinodal decomposition along the line. 
On the other hand, results from MPFCM-based 
modeling imply that the concentration 
dependence of the solute misfit distortion and resulting dependence of 
the elastic energy density on concentration have the strongest effect. The 
maximum amount of Mn segregating to straight edge dislocations predicted 
by MPFCM agrees well with APT results. On the other hand, the current MPFCM 
model for Fe-Mn predicts little or no variation in Mn concentration along 
a straight dislocation line, in contrast to the APT results. As shown by 
the example of a dislocation loop in the current work, a change in the 
hydrostatic stress along the line due to changing character of dislocation
does lead to a corresponding variation in Mn concentration. Such a variation 
in Mn concentration can also then be expected along a dislocation line with 
kinks or jogs.

\end{abstract} 


\end{frontmatter}

\section{Introduction}
The dependence of material properties on chemical composition, 
temperature and pressure (stress) is central to the phase relations, 
thermodynamics, and behavior, of many materials. In the case of 
engineering alloys, for example, the chemistry-dependence of 
elastic or magnetic material properties can have a profound 
influence on thermodynamics and heterogeneity. 
A classic example of this (also relevant to the current work) is 
the effect of lattice misfit on spinodal decomposition in 
such alloys as first discussed by Cahn \cite{Cahn1961,Cahn1962} 
and later by Khatchaturyan \cite{Kha83}. 

Even in the defect-free case, spatial variations in 
chemical composition for any reason result in heterogeneous material 
properties and behavior; the presence of defects further complicates this. 
Many experimental studies have documented the strong influence of 
defects on chemical heterogeneity in alloys. For example, in the case 
of FCC alloys (e.g., \cite{Kontis2018,Mianroodi2019a,Wu2020,LIU2021116966,Zhou2021}), 
such defects include dislocations, stacking faults, and grain boundaries and phase interfaces. 
Their respective influence on solute segregation and chemical 
heterogeneity differs in general. This is also the case in BCC materials 
such as the Fe-Mn alloys (e.g., \cite{Kuzmina2015,KwiatkowskiDaSilva2018a, kwiatkowski_da_silva_sustainable_2022}). 
Here, differences in Mn segregation behavior to grain boundaries 
\cite{KUZMINA2015182,KWIATKOWSKIDASILVA2019109} and to bulk dislocations 
\cite{Kuzmina2015,KWIATKOWSKIDASILVA2017305} have been experimentally 
observed and characterized, in particular with the help of atom probe 
tomography (APT). Such a dependence on defect type is related 
at least in part to the nature of the corresponding defect-solute 
interaction involved. In the case of grain boundaries or stacking faults, 
for example, this interaction is of short-range character; for dislocations 
on the other hand, it is clearly long-range in nature. Since the seminal 
work of Cottrell \cite{Cottrell49}, it has been known that the interaction 
of the (long-range) dislocation stress field with solute misfit results in 
strong solute segregation to such defects. More specifically, as shown by 
this and later work in the case of solute transport, or by that of Cahn 
\cite{Cahn1961} and later work in the case of spinodal decomposition, 
the chemistry dependence of solute lattice misfit results in a dependence 
of the solute chemical potential on the elastic energy and hydrostatic 
stress field. In turn, this results in a dependence of both the driving force 
for solute transport and spinodal decomposition on these. 

Building on this, more recent work has focused on further aspects and details 
of this type of chemoelastic coupling. For example, in \cite{Ma2006}, solute 
segregation and "wetting transition" at stationary and gliding dislocations 
has been modeled with the help of linear elastic phase-field microelasticity 
\cite[PFM:][]{Wan01} and the Cahn-Hilliard (CH) \cite{Cahn1958} model. They concluded 
that short-range chemical interaction among solute atoms lies behind a "wetting 
transition" at the core, depending in particular on temperature and solute 
concentration. Quite recently, \cite{BARKAR2018446} combined the linear elastic 
chemomechanical model of Cahn \cite{Cahn1961} with a dependence 
of the gradient chemical energy on solute concentration due to magnetic 
transitions to model defect-free spinodal decomposition in Fe-Cr. 

The purpose of the current work is comparative modeling and experimental 
investigation of segregation to, and spinodal decomposition at, dislocations 
in BCC Fe-Mn. On the modeling side, both microscopic phase-field chemomechanics 
(MPFCM) and (hybrid) Monte Carlo molecular dynamics (MCMD) \cite{Sadigh2012} 
are employed. The former represents an application of the phase-field-based 
methodology developed in \cite{Svendsen2018} for the finite-deformation 
chemomechanics of multiphase, multicomponent solids to the case of 
microscopic dislocation-solute interaction. MPFCM has been applied to model 
such interaction and processes in other alloys as well 
(e.g., Ni-superalloys like Ni-Al-Co in \cite{Mianroodi2019a}). As in the 
case of Fe-Cr mentioned above, the current application to Fe-Mn involves 
in particular a dependence of the gradient chemical energy on Mn concentration 
to account for magnetic transition in this system with increasing Mn content. 

The work is organized as follows. In Section \ref{sec:Mod}, the form 
of the MPFCM model relevant to Fe-Mn is briefly summarized. This is 
followed in Section \ref{sec:Ide} by identification / calibration of 
the MPFCM model with the help of an MEAM-based potential as well as 
Thermo-Calc / CALPHAD data for Fe-Mn. APT characterization results for 
Mn segregation to dislocations in two Fe-Mn alloys are presented in 
Section \ref{sec:ResExp}. After discussing simulation details in 
Section \ref{sec:DetSim}, the APT and simulation results from 
MPFCM and MCMD are compared and discussed in detail in Section 
\ref{sec:ResSim}. The work ends with a summary and discussion 
in Section \ref{sec:DisSum}. 

\section{Microscopic phase-field chemomechanics model for Fe-Mn}
\label{sec:Mod}

Restricting attention to quasi-static, isothermal (i.e., fixed absolute 
temperature \(\theta\)) conditions, an MPFCM model for Fe-Mn is based 
in particular on the chemomechanical balance relations
\begin{equation}
\skew3\dot{c}
=\mathop{\mathrm{div}}
M\,\nabla\mu
\,,\quad
\bm{0}
=\mathop{\mathrm{div}}
\bm{P}
\,,
\label{equ:BalMix}
\end{equation} 
for Mn mass and alloy linear momentum, respectively. 
These determine the Mn concentration field \(c=c_{\mathrm{Mn}}\) 
and the alloy deformation field \(\bm{\chi}\), respectively. 
In these relations, \(\mu:=\mu_{\mathrm{Mn}}-\mu_{\mathrm{Fe}}\) is 
the chemical potential of Mn relative to Fe per unit mass (SI units J kg${}^{-1}$), 
\(\bm{P}\) represents the first Piola-Kirchhoff (PK) stress tensor, and 
\(M(c;\theta)\) is the mobility of Mn (SI units kg m${}^{2}$ J${}^{-1}$s${}^{-1}$).  

Besides the field relations \eqref{equ:BalMix}, the current model is determined 
by the form
\begin{equation}
\psi(c,\bm{F},\nabla c;\theta,\phi)
=\psi_{\mathrm{ho}}(c,\bm{F};\theta,\phi)
+\psi_{\mathrm{gr}}(c,\nabla c;\theta)
\label{equ:DenEneFre}
\end{equation} 
for the free energy density \(\psi\) consisting of (chemically) homogeneous 
\begin{equation}
\psi_{\mathrm{ho}}(c,\bm{F};\theta,\phi)
=\psi_{\mathrm{el}}(c,\bm{F};\theta,\phi)+\psi_{\mathrm{ch}}(c;\theta)
\label{equ:DenEneHomFre}
\end{equation} 
and chemically inhomogeneous \(\psi_{\mathrm{gr}}\) parts. In turn, 
\(\psi_{\mathrm{ho}}\) is determined by elastic \(\psi_{\mathrm{el}}\) and 
chemical \(\psi_{\mathrm{ch}}\) contributions, with \(\bm{F}=\nabla\bm{\chi}\) 
the alloy deformation gradient. As indicated by the notation, \(\psi\) 
depends parametrically on \(\theta\) and the disregistry field \(\phi\) of 
a stationary dislocation. As discussed in more detail in what follows, 
\(\phi\) is treated as a non-evolving (stationary dislocation) in the current work. 
The elastic contribution 
\begin{equation}
\psi_{\mathrm{el}}(c,\bm{F};\theta,\phi)
=\tfrac{1}{2}
\,\bm{E}_{\mathrm{L}}(c,\bm{F};\theta,\phi)
\cdot
\ssans{C}(c;\theta)
\,\bm{E}_{\mathrm{L}}(c,\bm{F};\theta,\phi)
\label{equ:DenEneFreEla}
\end{equation} 
to \(\psi\) is modeled in harmonic form based on the elastic stiffness 
\(\ssans{C}\) and assuming that the lattice (Green) strain 
\(
\bm{E}_{\mathrm{L}}
:=\tfrac{1}{2}(\bm{F}_{\mathrm{L}}^{\mathrm{T}}\bm{F}_{\mathrm{L}}-\bm{I})
\) 
is "small", i.e., \(|\bm{E}_{\mathrm{L}}|\ll 1\). The lattice local deformation 
\(
\bm{F}_{\mathrm{L}}:=\bm{F}\bm{F}_{\mathrm{R}}^{-1}
\) 
is determined in the current model by a residual local deformation 
\(
\bm{F}_{\mathrm{R}}(c;\theta,\phi)
=\bm{F}_{\mathrm{M}}(c;\theta)\,\bm{F}_{\mathrm{D}}(\phi)
\) 
due to 
(i) misfit of Mn in the BCC Fe lattice 
\(
\bm{F}_{\mathrm{M}}(c;\theta)
=\exp\lbrack 
(c-c_{0})\,\bm{N}_{\mathrm{M}}(\theta)
\rbrack
\,\bm{F}_{\mathrm{M}}(c_{0};\theta)
\), 
with \(\bm{N}_{\mathrm{M}}(\theta)\) the corresponding (infinitesimal) 
distortion per unit concentration, 
and (ii) lattice slip due to a single dislocation 
\(
\bm{F}_{\mathrm{D}}(\phi)
=\bm{I}+\phi\,\bm{b}\otimes\bm{n}/d
\) 
with Burgers vector \(\bm{b}\), glide-plane normal \(\bm{n}\), 
and glide-plane spacing \(d\). 
The homogeneous part \(\psi_{\mathrm{ch}}(c;\theta)\) of the chemical 
contribution to \(\psi\) is determined directly using the 
ThermoCalc \cite{Andersson2002} database for Fe-Mn as explained in 
Section \ref{sec_chem}. In addition, the generalized CH 
form\footnote{Note that \(\kappa\) in 
\cite{Cahn1958} is based on energy per atom, whereas \(\kappa\) in 
\eqref{equ:DenEneFreGra} is based on energy density.}
\begin{equation}
\psi_{\mathrm{gr}}(c,\nabla c;\theta)
=\kappa(c;\theta)
\ \nabla c\cdot\nabla c
\label{equ:DenEneFreGra}
\end{equation}
of \(\psi_{\mathrm{gr}}\) for cubic symmetry is adopted. In contrast 
to the non-magnetic pair interaction case \cite{Cahn1958}, higher-order 
interaction and magnetic effects result here in a composition-dependent 
\(\kappa\) \cite{LIU1998222,BARKAR2018446}. 
Lastly, the above constitutive forms for \(\psi\) and \(\bm{F}_{\mathrm{R}}\) 
determine in turn those 
\begin{equation}
\bm{P}=\partial_{\bm{F}}\psi
\label{equ:StsKirPioFir}
\end{equation} 
for the first PK stress and 
\begin{equation}
\varrho\,\mu
=\tfrac{1}{2}
\,\bm{E}_{\mathrm{L}}
\cdot
(\partial_{c}\ssans{C})
\,\bm{E}_{\mathrm{L}}
-\bm{F}_{\mathrm{L}}
\bm{N}_{\mathrm{M}}
\bm{F}_{\smash{\mathrm{L}}}^{-1}
\cdot
\bm{K}
+\partial_{c}\psi_{\mathrm{ch}}
+(\partial_{c}\kappa)\,|\nabla c|^{2}
-\mathop{\mathrm{div}}2\,\kappa\,\nabla c
\,,
\label{equ:PotCheGenCH}
\end{equation}
for the CH chemical potential of Mn (\(\mu\equiv\mu_{\mathrm{Mn}}-\mu_{\mathrm{Fe}}\)). 
In this latter relation, \(\varrho\) is the mass density, and 
\(
\bm{K}
=\bm{P}\bm{F}^{\mathrm{T}}
\) 
is the Kirchhoff stress.
For more details on the current approach 
and methodology, the interested reader is referred to \cite{Svendsen2018}. 

\section{Model identification}
\label{sec:Ide}

The MPFCM model properties 
\(\ssans{C}(c;\theta)\), 
\(\bm{N}_{\mathrm{M}}(\theta)\), 
\(\phi\), 
\(M(c;\theta)\), and \(\kappa(c;\theta)\) are determined for Fe-Mn using 
the Version 9 database of Thermo-Calc Software TCFE Steels/Fe-alloys 
\cite{TCwebpage} and the modified embedded atom method (MEAM) potential 
for Fe-Mn from \cite{Kim2009a}. 

\subsection{Misfit distortion and elastic stiffness}
\label{sec_elas}

\begin{figure}[H]
\centering
\includegraphics[width=0.49\textwidth]{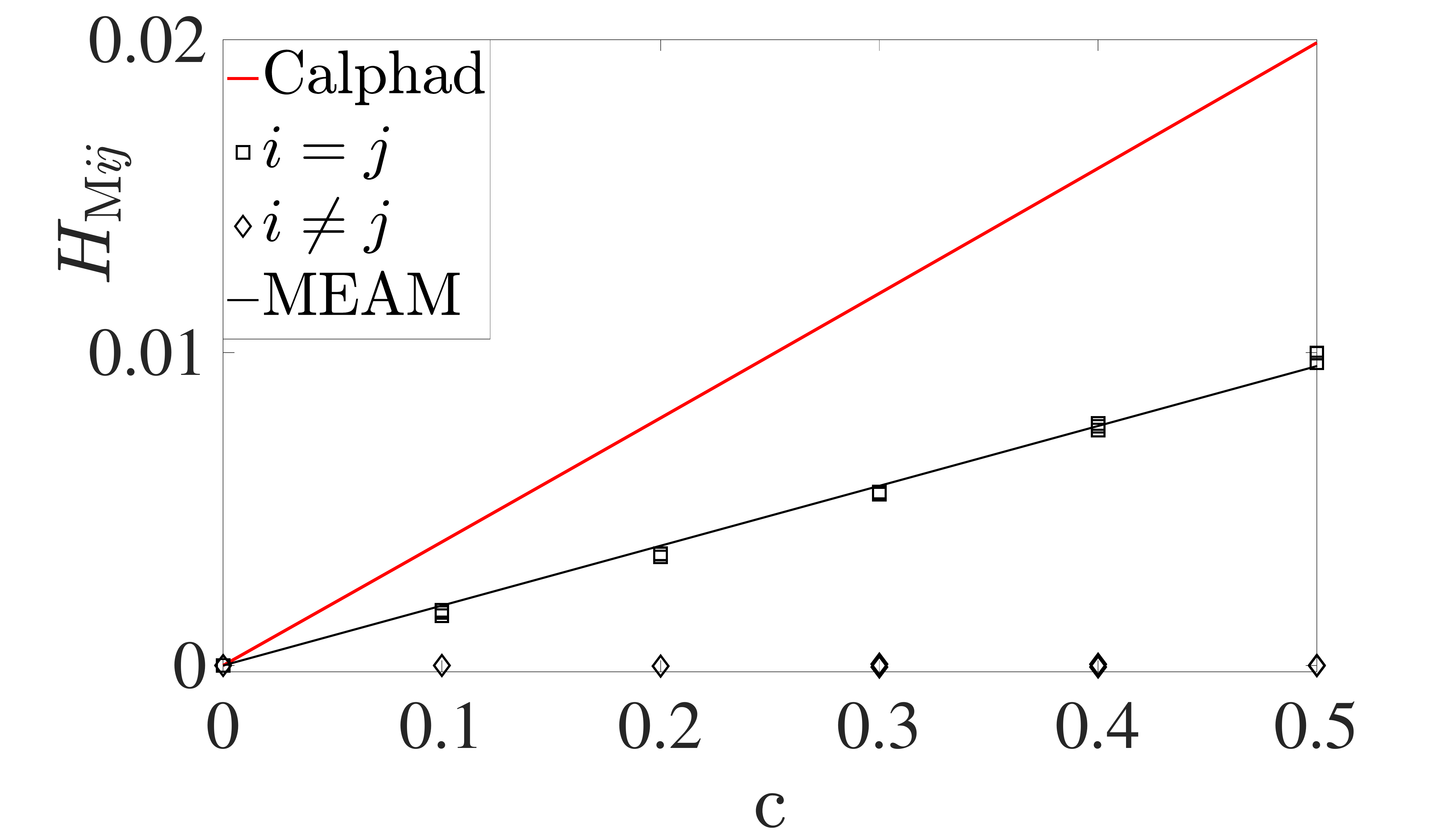}
\includegraphics[width=0.49\textwidth]{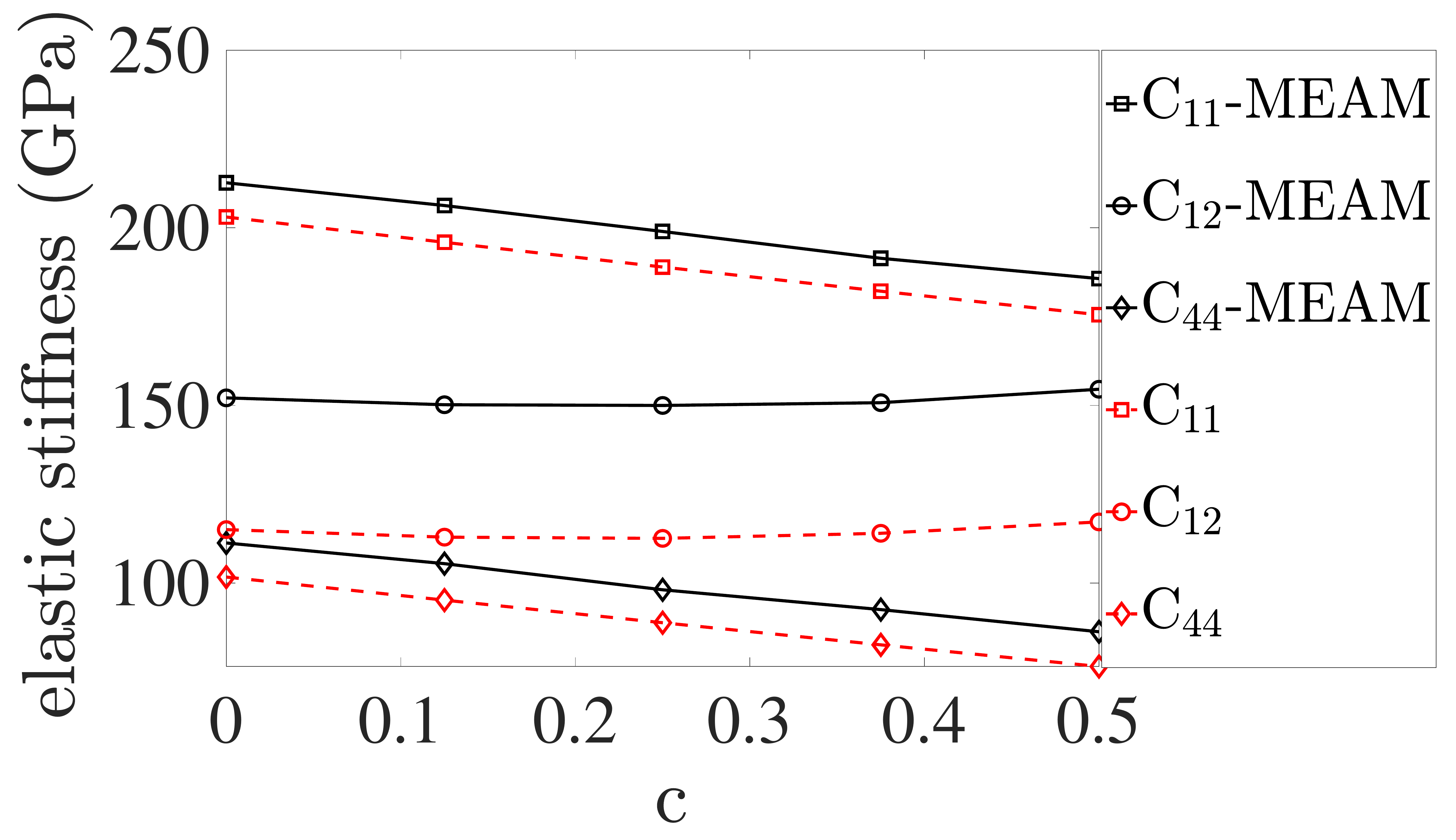}
\caption{Left:~MD simulation results (black squares, diamonds) 
for Mn misfit distortion in BCC Fe at 723 K and corresponding 
fit to \(H_{\mathrm{M}i\!j}(c;\theta)=c\,N_{\mathrm{M}i\!j}(\theta)\) 
with \(N_{\mathrm{M}i\!j}(\theta)=\nu_{\mathrm{M}}(\theta)\,\delta_{ij}\) 
and \(\nu_{\mathrm{M}}(723)=0.0194\) (black line). 
The red curve represents an analogous fit to Thermo-Calc data, yielding 
\(\nu_{\mathrm{M}}(723)=0.0398\). For both cases, 
\(c_{0}=0\) and \(\bm{F}_{\mathrm{M}}(c_{0};\theta)=\bm{I}\). 
The double black squares at \(c=0.4\) and \(c=0.5\) are due to small 
fluctuations in Mn distribution which have negligible effect on the fit. 
Right:~MD (black) and Thermo-Calc (red) results for the (cubic) 
components of \(\ssans{C}(c;723)\). Note that $c$ is in mole fraction. 
See text for details.}
\label{fig_elas}
\end{figure}

Determination of the local deformation 
\(
\bm{F}_{\mathrm{M}}(c;\theta)
\) 
due to Mn misfit is based on the corresponding distortion 
\(
\bm{H}_{\mathrm{M}}(c;\theta)
:=\bm{F}_{\mathrm{M}}(c;\theta)-\bm{I}
\). 
Results for this and for the elastic stiffness \(\ssans{C}(c;\theta)\) 
are shown in Figure \ref{fig_elas}. 
The MEAM results are based on MD simulations at 723 K. In these, 
Fe atoms are randomly replaced in the simulation cell with Mn atoms up 
to a given concentration. The cell is then relaxed to zero stress and 
the relevant change determined as a function of Mn concentration. 
As shown by the corresponding fit to the MEAM-based results, 
\(
\bm{H}_{\mathrm{M}}(c;\theta)
\) 
is well-approximated by 
\begin{equation}
\bm{H}_{\mathrm{M}}(c;\theta)
\approx
c\bm{N}_{\mathrm{M}}(\theta)
\,,\quad
\bm{N}_{\mathrm{M}}\theta)=\nu_{\mathrm{M}}(\theta)\,\bm{I}
\,,
\label{equ:MisLatSol}
\end{equation}
linear in \(c\) and geometrically linear. 
This is also true for the fit to Fe-Mn molar volume data in the Thermo-Calc 
database of CALPHAD (Figure \ref{fig_elas}, left, red curve). 
The \(c\)-dependence of the cubic components of \(\ssans{C}(c;723)\) 
shown in Figure \ref{fig_elas} (right) is determined in a similar fashion 
as that for the components of \(\bm{H}_{\mathrm{M}}(c;723)\) just discussed. 

\subsection{Chemical free energy}
\label{sec_chem}

\begin{figure}[H]
\centering
\includegraphics[width=0.49\textwidth]{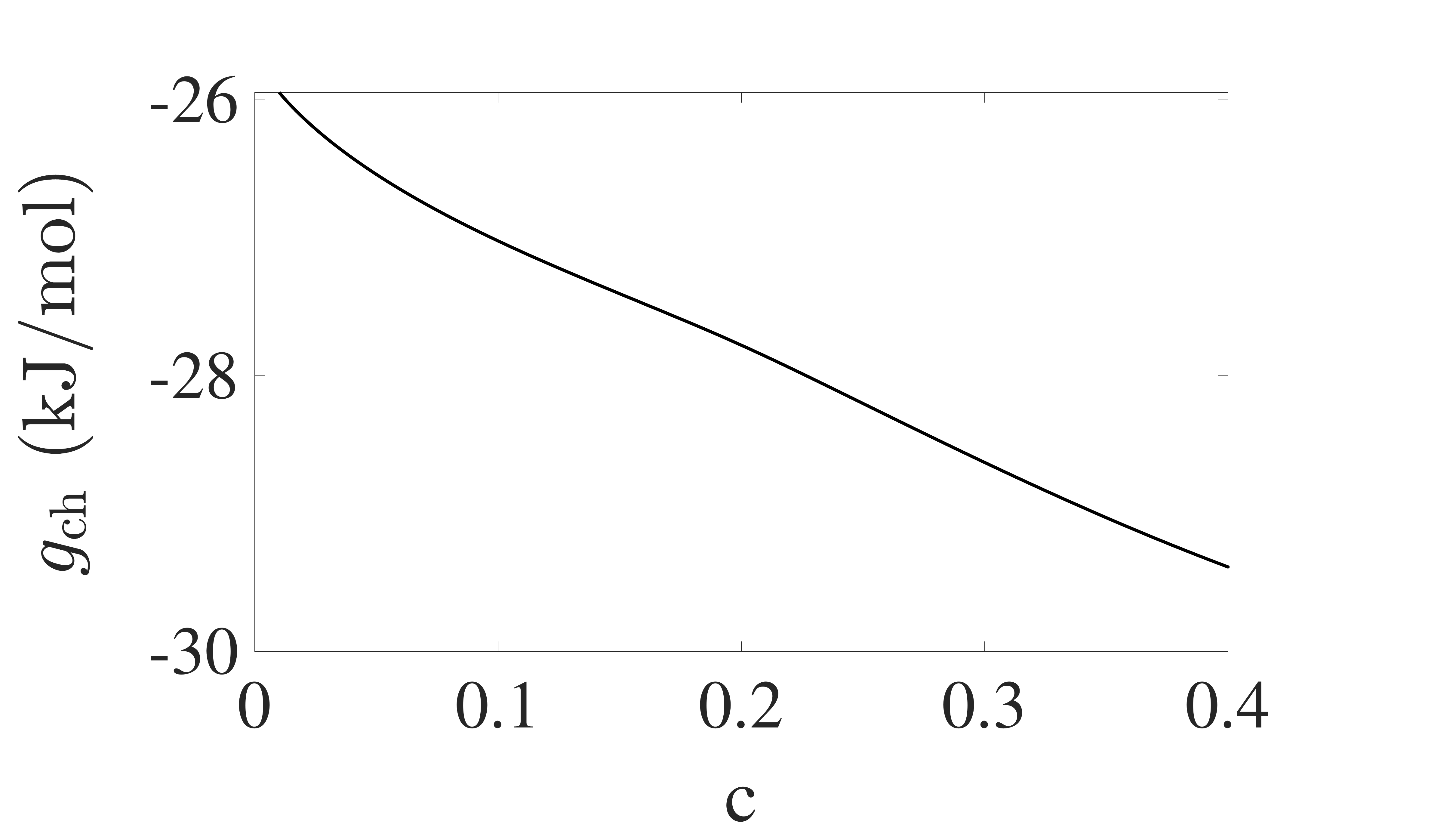}
\includegraphics[width=0.49\textwidth]{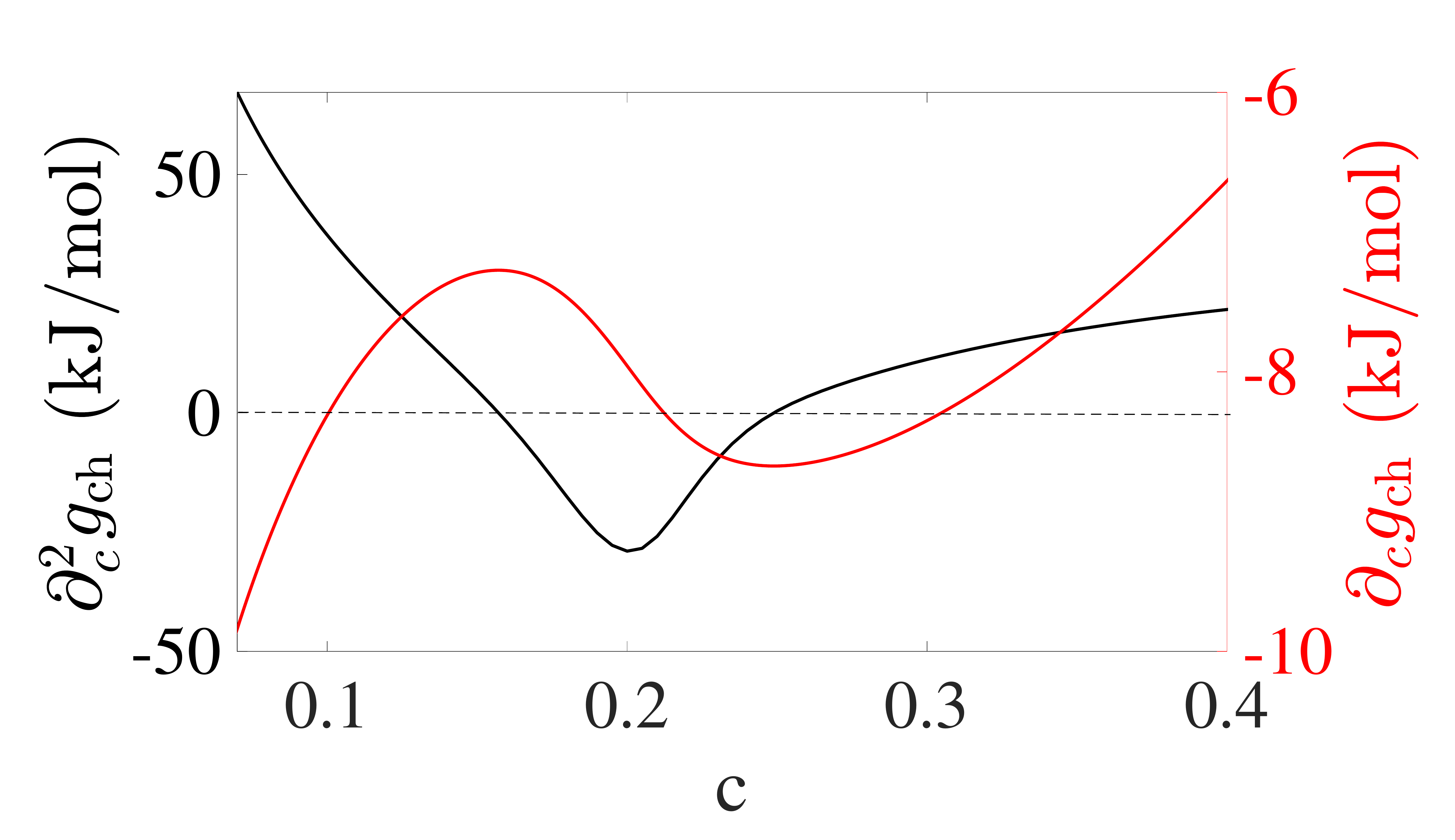}
\caption{Molar Gibbs free energy 
\(g_{\mathrm{ch}}(c;\theta=723)\) (left, black curve), 
\(\partial_{c}g_{\mathrm{ch}}(c;723)\) (right, red curve), and 
\(\partial_{c}^{2}g_{\mathrm{ch}}(c;723)\) (right, black curve), 
for Fe-Mn as a function of \(c\) from the Thermo-Calc database at 723 K. 
Note the spinodal points at \(c=0.157\) and \(c=0.249\).
}
\label{fig_freeEn}
\end{figure}

The homogeneous part 
\begin{equation}
\psi_{\mathrm{ch}}(c;\theta)
=g_{\mathrm{ch}}(c;\theta)/\Omega_{\mathrm{mol}}
\label{equ:DenEneFreCheHom}
\end{equation}
of the chemical free energy density in \eqref{equ:DenEneHomFre} is determined 
with the help of the molar volume \(\Omega_{\mathrm{mol}}\) and the 
corresponding molar Gibbs free energy \(g_{\mathrm{ch}}(c;\theta)\) 
from the Thermo-Calc database at room pressure and \(\theta=723\) K. This 
latter energy and its concentration derivatives are shown in 
Figure \ref{fig_freeEn}. 
A piecewise-spline fit to 200 discrete data segments exported from 
Thermo-Calc is used to determine \(g_{\mathrm{ch}}(c;723)\). In turn, 
this is used to determine \(\psi_{\mathrm{ch}}(c;\theta)\) using the 
constant value \(\Omega_{\mathrm{mol}}=7.45\times 10^{-6}\) m$^3$/mol 
for pure Fe at 723 K. 
The use of Thermo-Calc data ensures for example accurate spinodal points as 
well as correct account for magnetic effects at room pressure, not accounted 
for in the MEAM potential. 

\subsection{Gradient energy and mobility}
\label{sec_other}

\begin{figure}[H]
    \centering
    \includegraphics[width=0.49\textwidth]{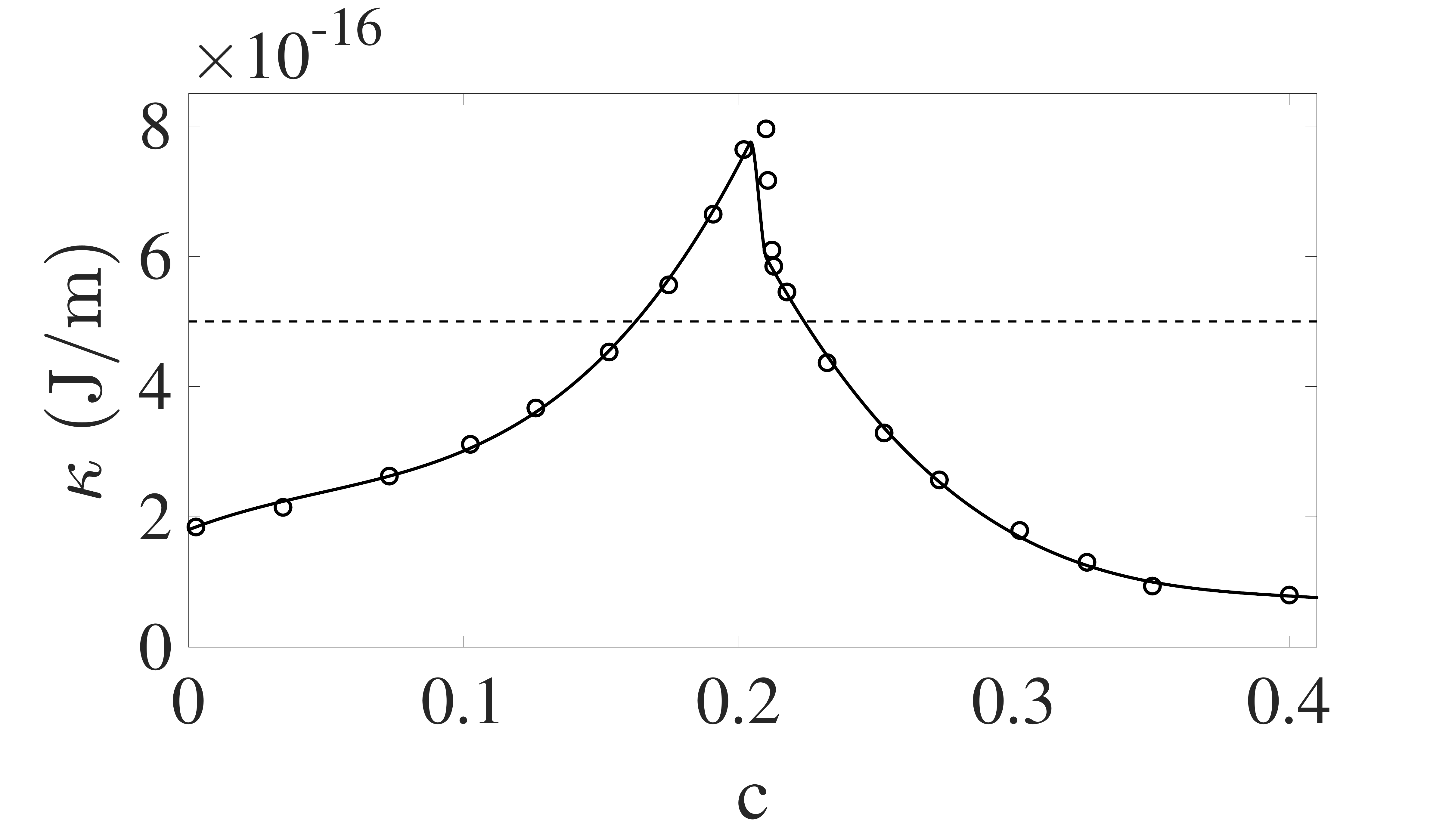}
    \includegraphics[width=0.49\textwidth]{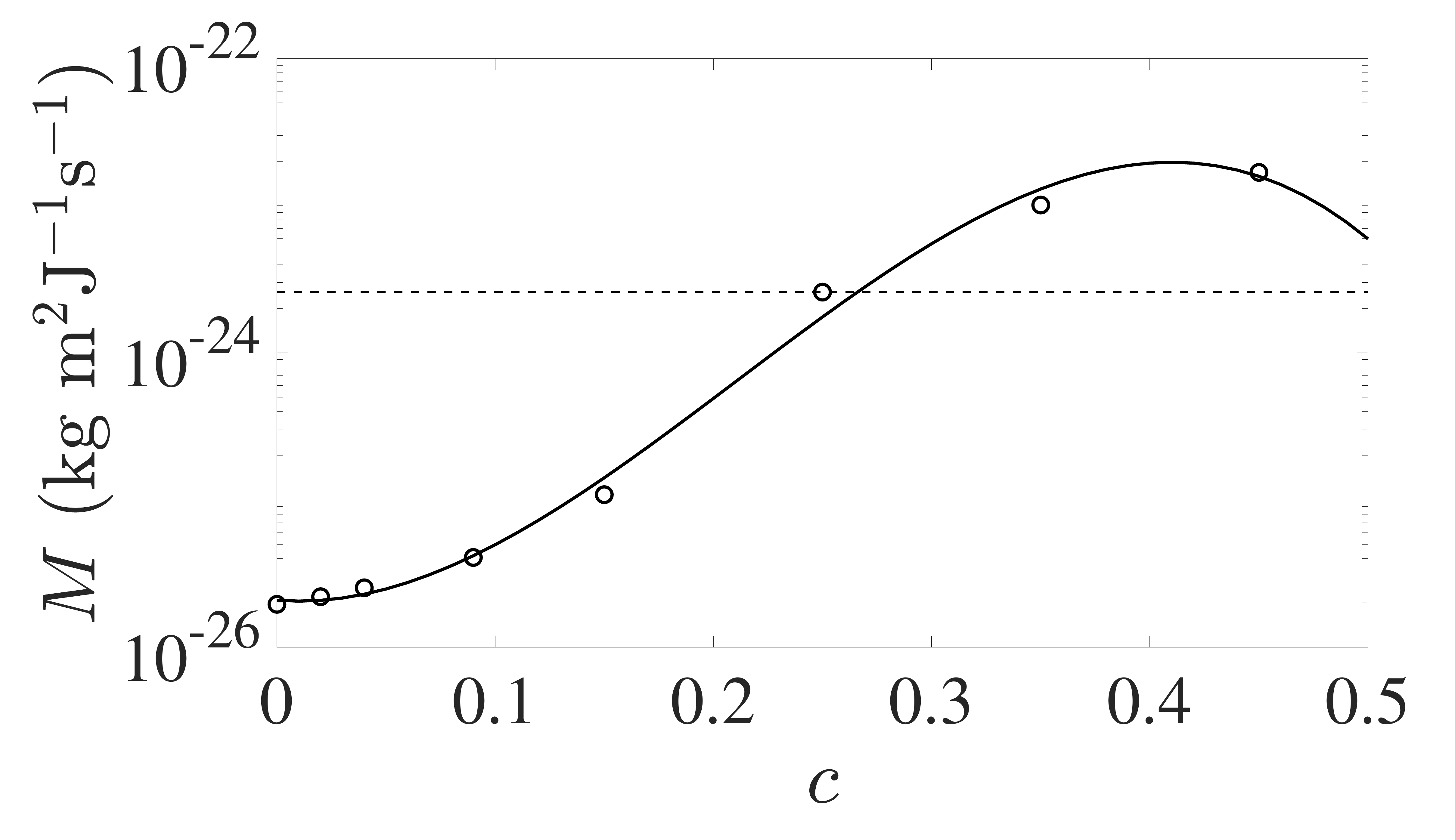}
    \caption{Results for the gradient energy coefficient \(\kappa(c;723)\) (left) 
    and mobility \(M(c;723)\) of Mn in BCC Fe (right). Shown are 
    data points (circles) from the database and the corresponding fits 
    for \(\kappa(c;723)\) (see \eqref{equ:CoeGraFitPol}) and 
    \(M(c;723)=M(0;723)\,\exp(-2.7185c-135.52c^2-214.94c^3)\)
    (solid curves). 
    Also shown are the constant values \(\kappa_{0}=5\times 10^{-16}\) J/m 
    (left, dotted line) and 
    \(M_{0}=M(0.25;723)=2.59\times 10^{-24}\) kg m${}^{2}$ J${}^{-1}$ s${}^{-1}$ 
    (right, dotted line) used in the simulations.}
    \label{fig_grad}
\end{figure}

Results for the gradient energy coefficient \(\kappa(c;\theta)\) 
and the mobility \(M(c;\theta)\) as a function of \(c\) at \(\theta=723\) K 
are shown in Figure \ref{fig_grad}. 
In particular, \(M(c;723)\) is obtained from Thermo-Calc mobility 
data \cite{TCwebpageMob} via polynomial fit. Following 
\cite{LIU1998222,BARKAR2018446}, Thermo-Calc data for \(\kappa(c;723)\) 
including magnetic contributions is fit to the piecewise continuous cubic 
polynomial 
\begin{equation}
\kappa(c;723)
=\left\lbrace
\begin{array}{lcl}
a_{0}+a_{1}\,c+a_{2}\,c^{2}+a_{3}\,c^{3}
&&
c<c_{0}
\\
a_{4}+a_{5}\,(c-c_{0})+a_{6}\,(c-c_{0})^{2}+a_{7}\,(c-c_{0})^{3}
&&
c_{0}\leqslant c<c_{1}
\\
a_{8}+a_{9}\,(c-c_{1})+a_{10}\,(c-c_{1})^{2}+a_{11}\,(c-c_{1})^{3}
&&
c\geqslant c_{1}
\end{array}
\,,
\right.
\label{equ:CoeGraFitPol}
\end{equation}
yielding \(c_{0}=0.204\), \(c_{1}=0.21\), and the values listed in Table \ref{tab:CoeGraFitPol}. 
\begin{table}[H]
\centering
\caption{Coeffients of \eqref{equ:CoeGraFitPol} fit to Thermo-Calc data for Fe-Mn at 723 K.
}
\begin{tabular}{|c|c|c|c|c|c|} \hline
\(a_{0}/10^{-16}\) & 
\(a_{1}/10^{-15}\) & 
\(a_{2}/10^{-14}\) & 
\(a_{3}/10^{-13}\) & 
\(a_{4}/10^{-16}\) & 
\(a_{5}/10^{-15}\) 
\\ \hline\hline
1.8 & 1.65 & -1.46 & 1.02 & 7.74 & 8.41 
\\ \hline
\end{tabular}

\vspace{2mm}

\begin{tabular}{|c|c|c|c|c|c|} \hline
\(a_{6}/10^{-11}\) & 
\(a_{7}/10^{-9}\) & 
\(a_{8}/10^{-16}\) & 
\(a_{9}/10^{-15}\) & 
\(a_{10}/10^{-14}\) & 
\(a_{11}/10^{-14}\)
\\ \hline\hline
-1.63 & 1.66 & 5.97 & -7.61 & 3.83 & -6.63
\\ \hline
\end{tabular}
\label{tab:CoeGraFitPol}
\end{table}
The peak and transition in 
\(\kappa(c;723)\) at \(c=c_{0}\) evident in Figure \ref{fig_grad} 
(left) is due to a transition from ferro- to paramagnetic 
(e.g., analogous to Fe-Cr:~\cite{BARKAR2018446}). 

\subsection{Dislocation disregistry field}

\begin{figure}[H]
    \centering
    \includegraphics[width=0.49\textwidth]{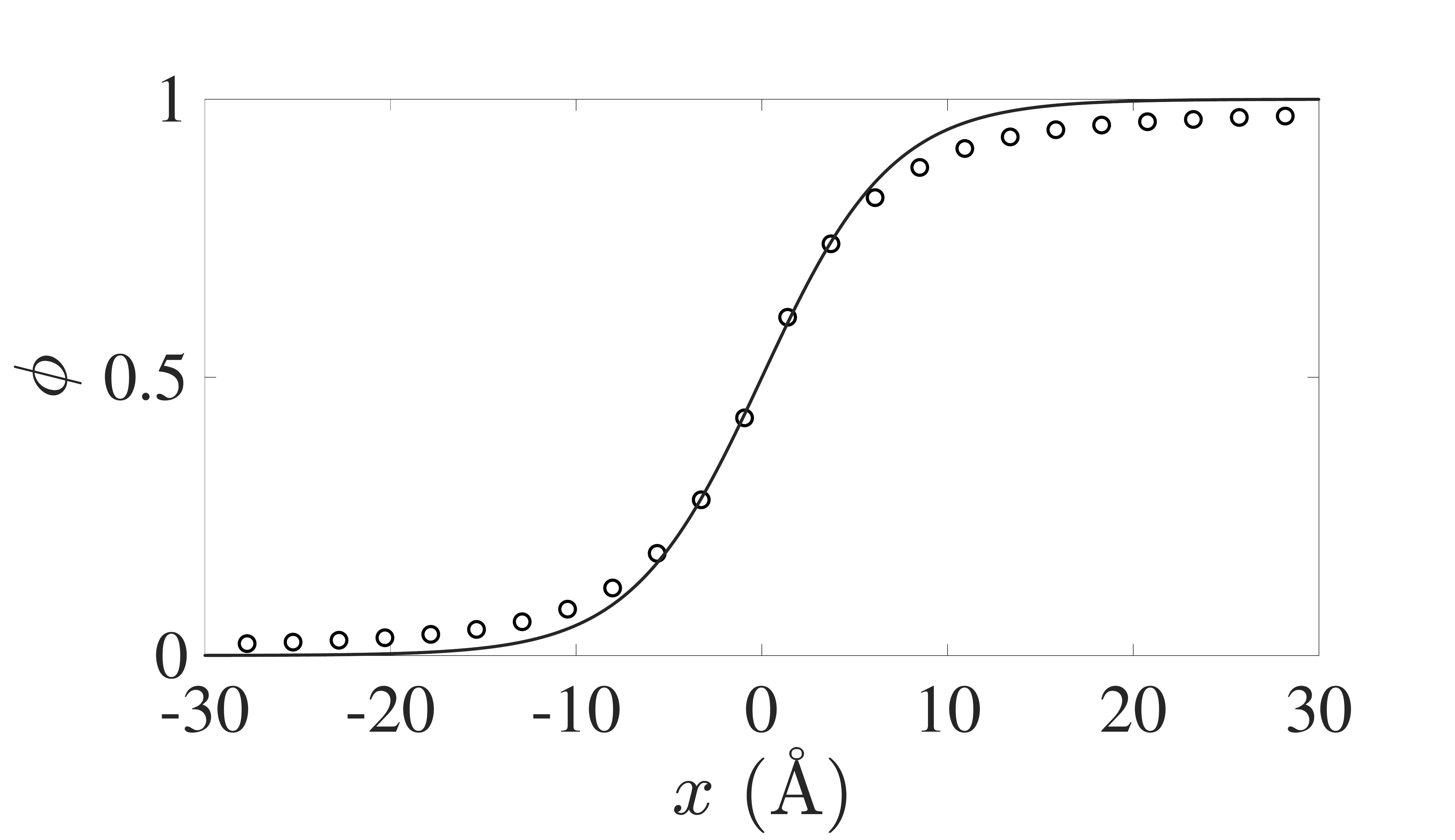}
    \caption{Arctan-based disregistry profile \(\phi=u_{x}/b_{\lbrack 111\rbrack}\) 
    (solid line) fit to the atomistic results (circles) for a perfect 
    \(\frac{1}{2}\langle 111\rangle\lbrace 110\rbrace\) edge dislocation 
    in BCC Fe obtained from molecular statics based on the MEAM potential at 0 K.\newline
    }
    \label{fig_core}
\end{figure}

As explained in Section \ref{sec:Mod}, the (fixed) dislocation disregistry 
field \(\phi=u/b\) is treated as a material property in the current work 
and determined using the Fe-Mn MEAM potential. The case of a perfect 
$\frac{1}{2}\langle 111\rangle\lbrace 110\rbrace$ edge dislocation in 
BCC Fe is shown in Figure \ref{fig_core}.
The results in Figure \ref{fig_core} determine \(\phi(0;0)\). Using the 
MEAM potential, the dependence of \(\phi\) on \(c\) was studied and found 
to be minimal. This is assumed in the following for the \(\theta\)-dependence 
as well. On this basis, all required material parameters have been identified 
for Fe-Mn in the current MPFCM model.

\section{Characterization of Mn segregation via atom probe tomography}
\label{sec:ResExp}

To investigate the effect of alloy composition on solute segregation to 
dislocations via atom probe tomography (APT), two Fe-Mn alloys 
(Fe7Mn and Fe9Mn) were prepared. 
The alloys were cast into rectangular billets in a vacuum induction furnace, 
homogenized at temperatures between 1100-1150 C, hot-rolled and quenched to room 
temperature, resulting in a martensitic microstructure. Both alloys were 
subsequently annealed at 723 K for 6 hours to facilitate Mn segregation 
to dislocations. The composition of the alloys is shown in Table \ref{tab_Alloy}. 
\begin{table}[H]
\footnotesize
\centering
\caption{Chemical composition of Fe-7wt\%Mn (Fe7Mn) and Fe-9wt\%Mn (Fe9Mn) 
based on wet chemical analysis.}
\begin{tabular}{|c|c|c|c|c|c|c|} \hline
      & Mn & C & Si & Al & P & S\\ \hline
Fe7Mn & 7.22 & 0.093 & 0.49 & 0.013 & 0.005 & 0.007  \\ \hline
Fe9Mn & 8.46 & 0.0075 & 0.0024 & $<$0.002 & $<$0.002 & 0.0047 \\ \hline
\end{tabular}
\label{tab_Alloy}
\end{table}
Note that Fe7Mn contains close to 0.1wt.\% C, which could strongly affect Mn 
segregation to dislocations. After 6 hours of annealing at 450 C, 
however, carbide precipitation \cite{KWIATKOWSKIDASILVA2018165} substantially 
depletes the amount of C in the matrix. For further details on these alloys, 
the reader is referred to \cite{Kuzmina2015, KwiatkowskiDaSilva2018a, 
KUZMINA2015182, HAN20161, HAN2017199, KWIATKOWSKIDASILVA2017305, 
KWIATKOWSKIDASILVA2019109}. 

\begin{figure}[H]
\centering
\includegraphics[width=0.55\textwidth]{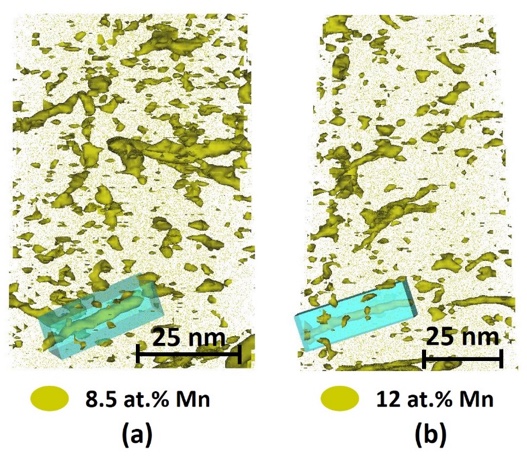}
\caption{APT reconstruction results for 
(a) Fe7Mn, 55\% cold-rolled and subsequently annealed at 723 K for 6 hours, 
and (b) Fe9Mn, 50\% cold-rolled and subsequently annealed at 723 K for 6 hours. 
Shown are 8.5 (left) and 12 (right) at.\% Mn isosurfaces revealing Mn segregation 
to dislocations. Dislocations in the blue regions were selected for further 
detailed compositional analysis.} 
\label{fig_APT}
\end{figure}

\begin{figure}[H]
\centering
\includegraphics[width=0.7\textwidth]{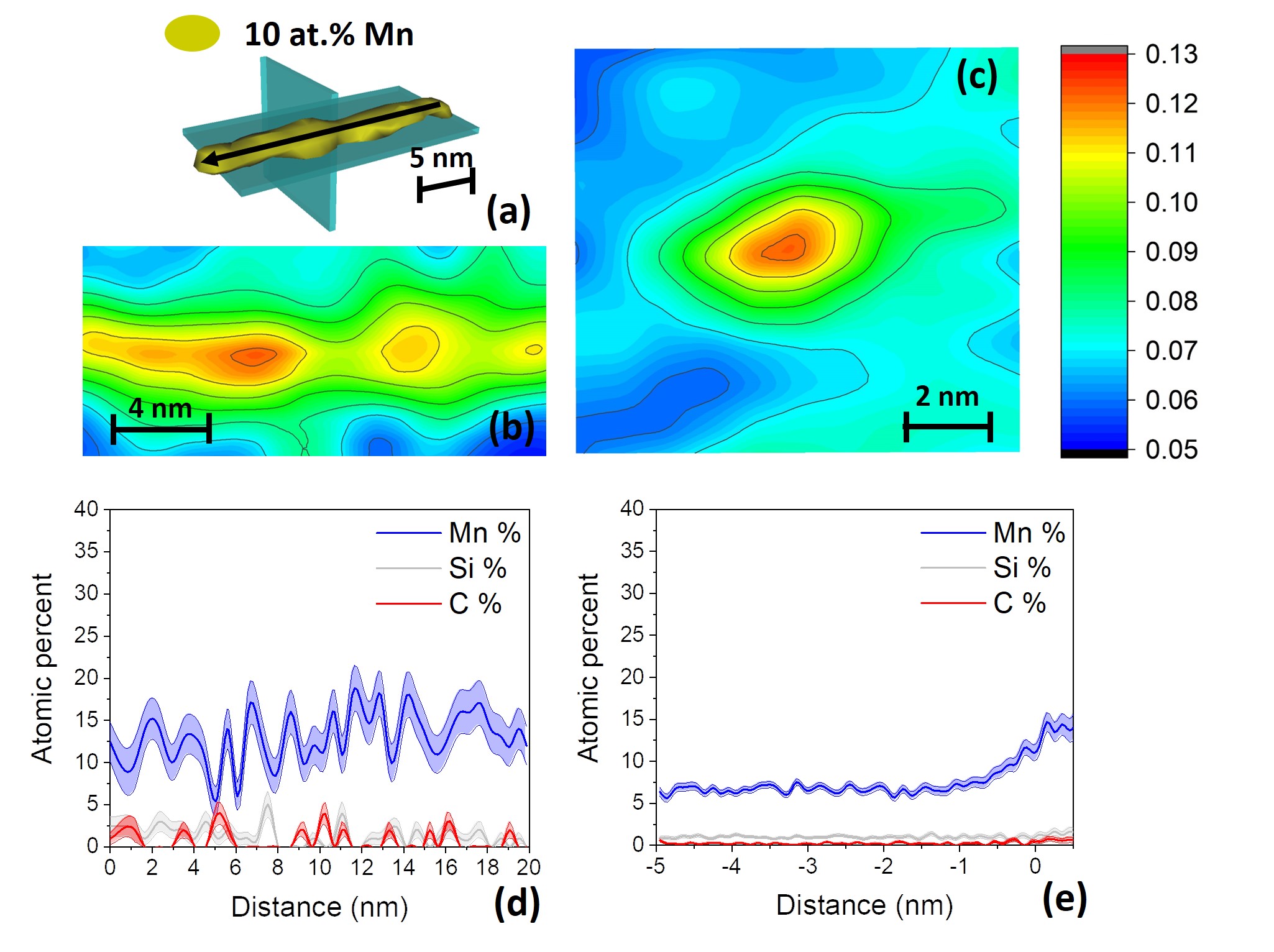}
\caption{APT results for segregated Mn at a dislocation 
(Figure \ref{fig_APT}a, blue region) in Fe7Mn. 
(a) 10 at.\% Mn isosurface, planes parallel and perpendicular to line. 
(b) at.\% Mn in plane parallel to the line. 
(c) at.\% Mn in plane perpendicular to the line. 
(d) at.\% Mn (violet), C (red), Si (gray) along the line. 
(e) at.\% Mn (violet), C (red), Si (gray) vs. distance perpendicular to the 
10 at.\% Mn isosurface in (a). The bands around the curves in 
(d) and (e) represent APT measurement uncertainty. } 
\label{fig_APT_7}
\end{figure}

\begin{figure}[H]
\centering
\includegraphics[width=0.7\textwidth]{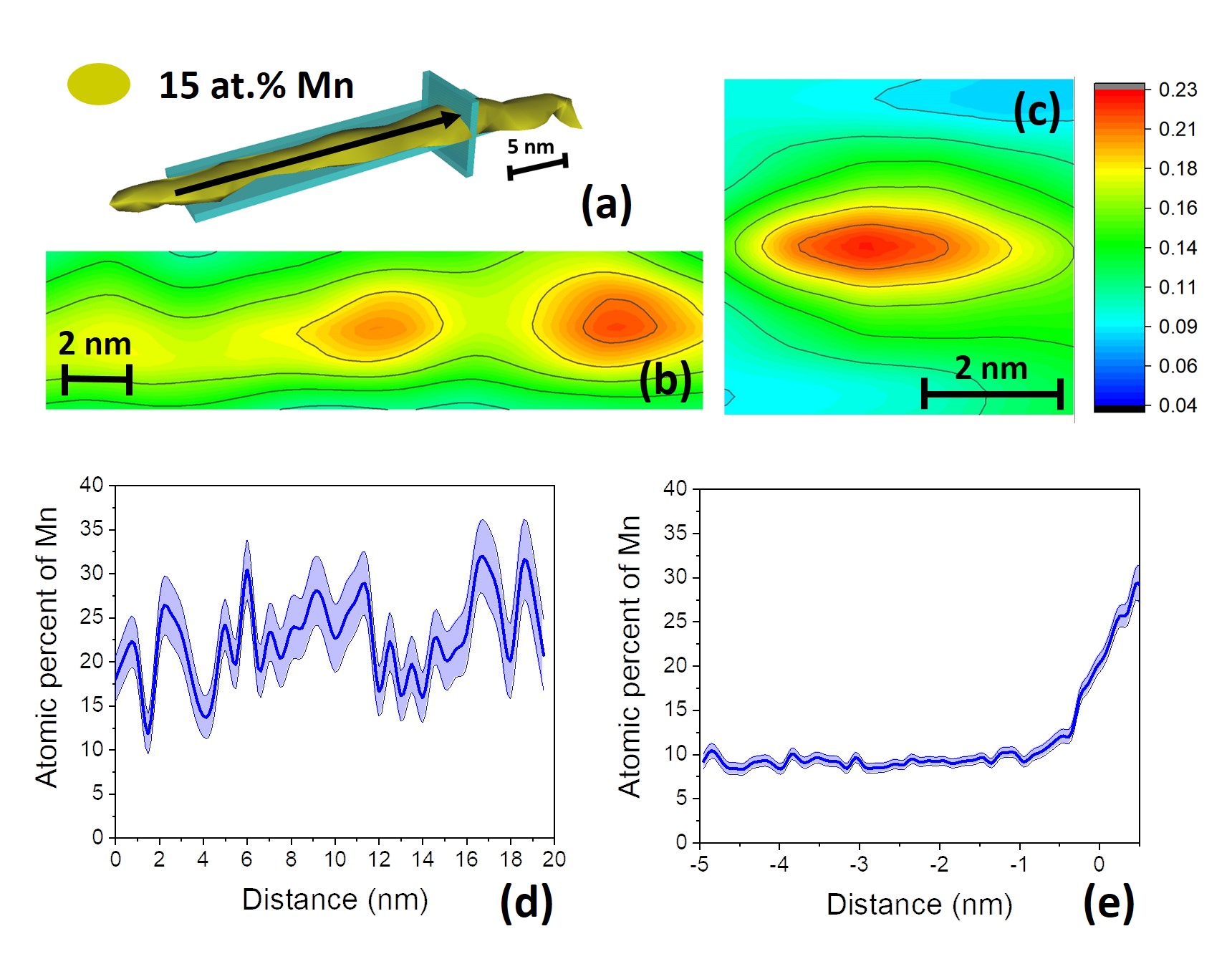}
\caption{APT results for segregated Mn at dislocation 
(Figure \ref{fig_APT}b, blue region) in Fe9Mn. 
See caption of Figure \ref{fig_APT_7} for details.} 
\label{fig_APT_9}
\end{figure}

For characterization via atom probe tomography (APT), specimens of these alloys 
with end radii below 100nm were prepared using a FEI Helios NanoLab600i dual-beam 
Focused Ion Beam (FIB) / Scanning Electron Microscopy (SEM) instrument. APT was 
performed using a Cameca Scientific Instruments (CIS) LEAP 5000 XS device. 
Data was obtained with this instrument 
(i) at approximately 80\% detection efficiency, 
(ii) at a set-point temperature of 50 K (Fe7Mn) and 60 K (Fe9Mn), 
(iii) in laser-pulsing mode at 355 nm and a pulsing rate of 500 kHz, and 
(iv) at a pulse energy of 40 pJ (Fe7Mn) and 30 pJ (Fe9Mn). 
For (re)construction of 3D atom maps from the data, as well as the visualization 
and quantification of segregation, the CIS software 
IVAS$^{\hbox{\tiny{\textregistered}}}$ was employed following the protocol 
introduced in \cite{Geiser2009} and detailed in \cite{GAULT2011448}. The 3D 
atom maps were obtained by voltage-based reconstruction of the detected 
ions. Reconstructions were calibrated by the interplanar distance of the 
crystallographic planes associated with the low-hit density poles. 
Figure \ref{fig_APT} displays resulting reconstructions for Fe7Mn and Fe9Mn. 
The lenticular distribution of Mn at.\% isosurfaces can be interpreted as 
segregation to dislocation lines. Identification of these as edge dislocations 
was carried out in previous work via correlated transmission electron 
microscopy / APT together with atom probe crystallographic analysis 
\cite{Kuzmina2015,KWIATKOWSKIDASILVA2017305}. 
Additional detailed composition analysis has been carried out on the 
dislocations inside the blue regions in Figure \ref{fig_APT}. The results 
of this for Fe7Mn are shown in Figure \ref{fig_APT_7}, and for Fe9Mn in 
Figure \ref{fig_APT_9}. 
\vspace{-5mm}
All at.\% results from APT reported here are based on an averaging volume 
of 0.125 nm${}^{3}$. As indicated by these results, the maximum amount of 
Mn segregated to (bulk edge) dislocations increases from about 16 at.\% in 
Fe7Mn to about 30 at.\% in Fe9Mn. In addition, the amount of segregated Mn 
fluctuates strongly along the line (i.e., between 5-16 at.\% in Fe7Mn and 
15-30 at.\% in Fe9Mn), in agreement with previous observations 
\cite{KwiatkowskiDaSilva2018a}. Note also the fluctuation of at.\% Si and 
C along the line (Figure \ref{fig_APT_7}d) approximately asynchronous to 
that of Mn. 

\section{Simulation details}
\label{sec:DetSim}

\subsection{Numerical solution of initial-boundary-value problems based on MPFCM}

This employs in particular the "weak" form 
\cite{Ubachs2004,Shanthraj2020} 
\begin{equation}
\psi_{\mathrm{wgr}}(c,\breve{c},\nabla\breve{c};\theta,\phi)
:=\tfrac{1}{2}\,\alpha\,(c-\breve{c})^{2}
+\psi_{\mathrm{gr}}(\breve{c},\nabla\breve{c};\theta)
\label{equ:DenEneFreCHWea}
\end{equation}
of the gradient energy \eqref{equ:DenEneFreGra} 
in terms of the auxiliary field \(\breve{c}\) and penalty parameter 
\(\alpha\). In this context, the difference between \(\breve{c}\) and 
\(c\) is minimized via minimization of the last two terms 
in \eqref{equ:DenEneFreCHWea} with respect to \(\breve{c}\). 
As usual, the corresponding Euler-Lagrange relation 
\begin{equation}
\delta_{\breve{c}}\psi_{\mathrm{wgr}}
=\alpha\,(\breve{c}-c)
+(\partial_{\breve{c}}\kappa)\,|\nabla\breve{c}|^{2}
-\mathop{\mathrm{div}}2\,\kappa\,\nabla\breve{c}
=0
\label{equ:PotCheCHWea0}
\end{equation} 
is necessary for this and provides a field relation for \(\breve{c}\). 
In the context of \eqref{equ:DenEneFreCHWea}, 
\begin{equation}
\varrho\,\mu
=\tfrac{1}{2}
\,\bm{E}_{\mathrm{L}}
\cdot
(\partial_{c}\ssans{C})
\,\bm{E}_{\mathrm{L}}
-\bm{F}_{\mathrm{L}}\bm{N}_{\mathrm{M}}\bm{F}_{\smash{\mathrm{L}}}^{-1}
\cdot
\bm{K}
+\partial_{c}\psi_{\mathrm{ch}}
+\alpha\,(c-\breve{c})
\label{equ:PotCheCHWea}
\end{equation} 
holds for \(\mu\) instead of \eqref{equ:PotCheGenCH}. 
Numerical solution of the field relations \eqref{equ:BalMix} and 
\eqref{equ:PotCheCHWea0} is 
carried out in a staggered fashion. Initial conditions here include a 
uniform Mn concentration in each case. Boundary conditions include zero 
external loading (stress control). Iteration proceeds until \(\dot{c}=0\). 
A value of $\alpha=10^{20}$ J/m${}^{3}$ for the penalty parameter is employed in all simulations. Changing 
this value to $10^{19}$ or $10^{21}$ has no influence on the simulation results. 

\subsection{Monte Carlo molecular dynamics}

For further comparison, Monte Carlo molecular 
dynamics (MCMD:~e.g., \cite{Sadigh2012}) simulations based on the MEAM potential 
are also performed. In the current application, each MCMD step consists of 
5000 MC steps (with 100 swaps per step) at 723 K followed by 5000 MD steps 
in the Parinello-Rahmann (NPT) ensemble at zero stress and 723 K. Such 
MCMD steps are repeated until the system and site occupancy stop changing. 
Averaging over the corresponding Mn lattice site occupation ensembles 
then yields the site Mn at.\% at zero (simulation cell) stress and 723 K. 

It is worth emphasizing that a number of differences exist between MCMD 
and MPFCM modeling of segregation behavior. These include in particular 
the energy models and stress measures. An indication of the former is 
the absence of screw core spreading in the MPFCM core stress field results 
in Figure \ref{fig_PFvsMCMD} (lower middle right) seen in the corresponding 
MCMD results in Figure \ref{fig_PFvsMCMD} (lower right). In the latter case, 
the atomic (virial) "stress" in MCMD is not a continuum stress measure. 
In addition, the magnitude of the atomic stress is inversely proportional 
to the atomic "volume" as based on Voronoi tessellation. Since this volume is physically arbitrary, the absolute value of the atomic stress 
is also physically arbitrary. Consequently, the comparison of MCMD and 
MPFCM results in what follows is at best qualitative in character. 

\subsection{Simulation set-up}

Unless otherwise stated, all simulation cells are fully periodic and Cartesian 
with cell side vectors 
\(L_{x}\bm{i}_{x}\), 
\(L_{y}\bm{i}_{y}\), 
\(L_{z}\bm{i}_{z}\). 
In the case of edge dislocations, for example, 
\begin{equation}
(L_x,L_y,L_z) = (60\sqrt{3},40\sqrt{2},2\sqrt{6})\,a_0
\,,\ 
(\bm{i}_{x},\bm{i}_{y},\bm{i}_{z})
=(\tfrac{1}{\sqrt{3}}\lbrack 111\rbrack,
\tfrac{1}{\sqrt{2}}\lbrack \bar{1}10\rbrack,
\tfrac{1}{\sqrt{6}}\lbrack 11\bar{2}\rbrack)
\,.
\label{equ:SimDisEdg}
\end{equation}
Analogously, 
\begin{equation}
(L_x,L_y,L_z) = (43\sqrt{6},40\sqrt{2},3\sqrt{3})\,a_0
\,,\ 
(\bm{i}_{x},\bm{i}_{y},\bm{i}_{z})
=(\tfrac{1}{\sqrt{6}}\lbrack 11\bar{2}\rbrack,
\tfrac{1}{\sqrt{2}}\lbrack \bar{1}10\rbrack,
\tfrac{1}{\sqrt{3}}\lbrack 111\rbrack)
\label{equ:SimDisScr}
\end{equation}
for screw dislocations. Here, \(a_0\) is the lattice constant of BCC 
Fe (2.586 \AA\ at 0 K). Simulations are based on a perfect 
\(\tfrac{1}{2}\langle 111\rangle\lbrace 110\rbrace\) 
dislocation dipole configuration with 
glide plane normal \(\bm{i}_{y}\) and monopole separation \(\frac{1}{2}L_x\). 
System size is chosen large enough to avoid any size dependency in the results. 

\section{Simulation results and comparison with APT data}
\label{sec:ResSim}

\subsection{Mn segregation to edge and screw dislocations}

For brevity, "APFCM" (atomistic PFCM) refers in what follows 
to the MPFCM model for Fe-Mn calibrated using MEAM-MD-based results for 
\(\bm{F}_{\mathrm{M}}\) and \(\ssans{C}\). Analogously, "MPFCM" refers 
to this model calibrated using Thermo-Calc data from CALPHAD 
(Figure \ref{fig_elas}). 

\begin{figure}[H]
\centering
\includegraphics[width=0.85\textwidth]{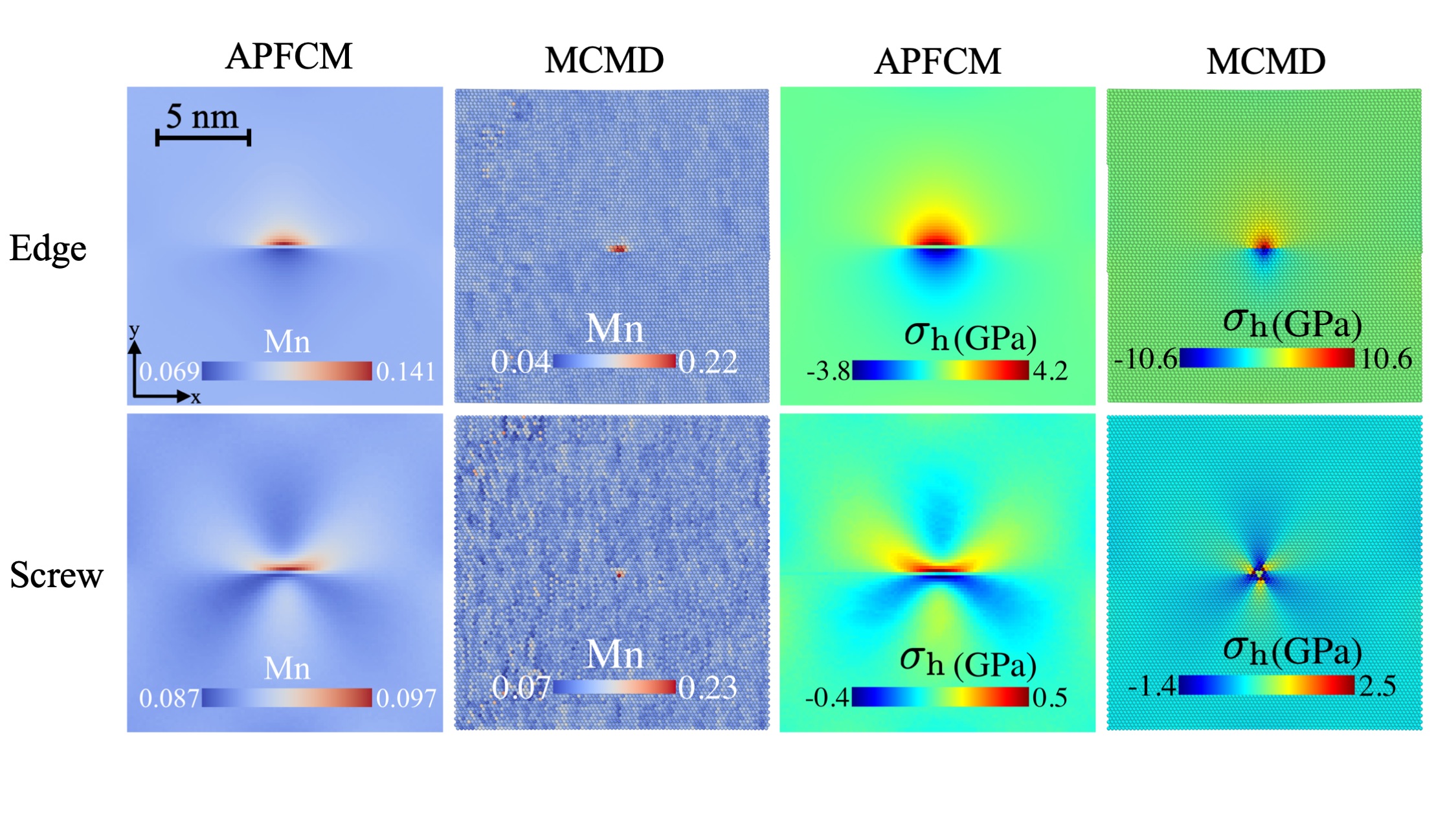}
\caption{Segregated Mn concentration (left) and 
hydrostatic stress field \(\sigma_{\mathrm{h}}\) (right) 
associated with $\frac{1}{2}\langle 111\rangle\lbrace 110\rbrace$ 
edge (top) and screw (bottom) dislocations predicted at 723 K 
by APFCM and MCMD.  In all cases, the average Mn concentration is 9 at.\%, 
corresponding to the experimental Fe9Mn case in Figure \ref{fig_APT_9}. 
Note that \(\sigma_{\mathrm{h}}\) is determined in APFCM by 
\(
\frac{1}{3}\,\bm{I}\cdot\bm{K}
=\frac{1}{3}\,K_{ii}
\), 
and in MCMD, by one-third the trace of the atomic stress. 
See text for discussion.} 
\label{fig_PFvsMCMD}
\end{figure}

APFCM and MCMD results for Mn segregation to 
$\frac{1}{2}\langle 111\rangle\lbrace 110\rbrace$ 
edge and screw dislocations are presented in Figure \ref{fig_PFvsMCMD}. 

As evident in \eqref{equ:PotCheGenCH}, the driving force \(\nabla\mu\) for 
solute flux in \eqref{equ:BalMix}${}_{1}$ depends in particular on (the gradient of) 
Kirchhoff stress \(\bm{K}\). In particular, since the identified form 
\eqref{equ:MisLatSol} of \(\bm{H}_{\mathrm{M}}(c;\theta)\) is dilatational, 
the model relation \eqref{equ:PotCheGenCH} for the chemical potential simplifies 
to the form 
\begin{equation}
\varrho\,\mu
=\tfrac{1}{2}
\,\bm{E}_{\mathrm{L}}
\cdot
(\partial_{c}\ssans{C})
\,\bm{E}_{\mathrm{L}}
-\nu_{\mathrm{M}}\,\bm{I}\cdot\bm{K}
+\partial_{c}\psi_{\mathrm{ch}}
+(\partial_{c}\kappa)\,|\nabla c|^{2}
-\mathop{\mathrm{div}}2\,\kappa\,\nabla c
\label{equ:PotCheRedCH}
\end{equation} 
depending only on the hydrostatic part 
\(\frac{1}{3}\bm{I}\cdot\bm{K}=\frac{1}{3}\,K_{ii}\) of \(\bm{K}\). Besides 
in the edge core (Figure \ref{fig_PFvsMCMD}, upper right) as expected, note 
that both APFCM (Figure \ref{fig_PFvsMCMD}, lower middle right) 
and MCMD (Figure \ref{fig_PFvsMCMD}, lower right) predict non-zero 
hydrostatic stress in the screw core, resulting in segregation to screws 
as well. This is in contrast to standard (i.e., Volterra) continuum 
dislocation theory (e.g., \cite[][]{Hirth1982}) based on linear elasticity 
and due in particular to geometric non-linearity. 
Additional non-linear effects accounted for in MCMD such as screw core 
spreading or core Kanzaki forces \cite{GLerma2018} may also play a role here. 
Especially in the APFCM case, the amount of segregation is clearly correlated 
with the magnitude of the (positive) hydrostatic stress field. 
The MCMD segregation results in Figure \ref{fig_PFvsMCMD} (middle left) 
hold for spatial regions smaller than can be resolved with APT. Averaging 
these "raw" MCMD values over the same volume ($0.125$ nm$^3$) used to obtain 
the APT results reduces for example the peak edge value of 22 at.\% Mn to 
about 17 at.\%, and the peak screw value of 23 at.\% to about 16 at.\%, both 
closer to the corresponding APFCM values. 

\subsection{Influence of elasticity and defects on 
segregation and spinodal behavior}

\begin{figure}[H]
\centering
\includegraphics[width=0.7\textwidth]{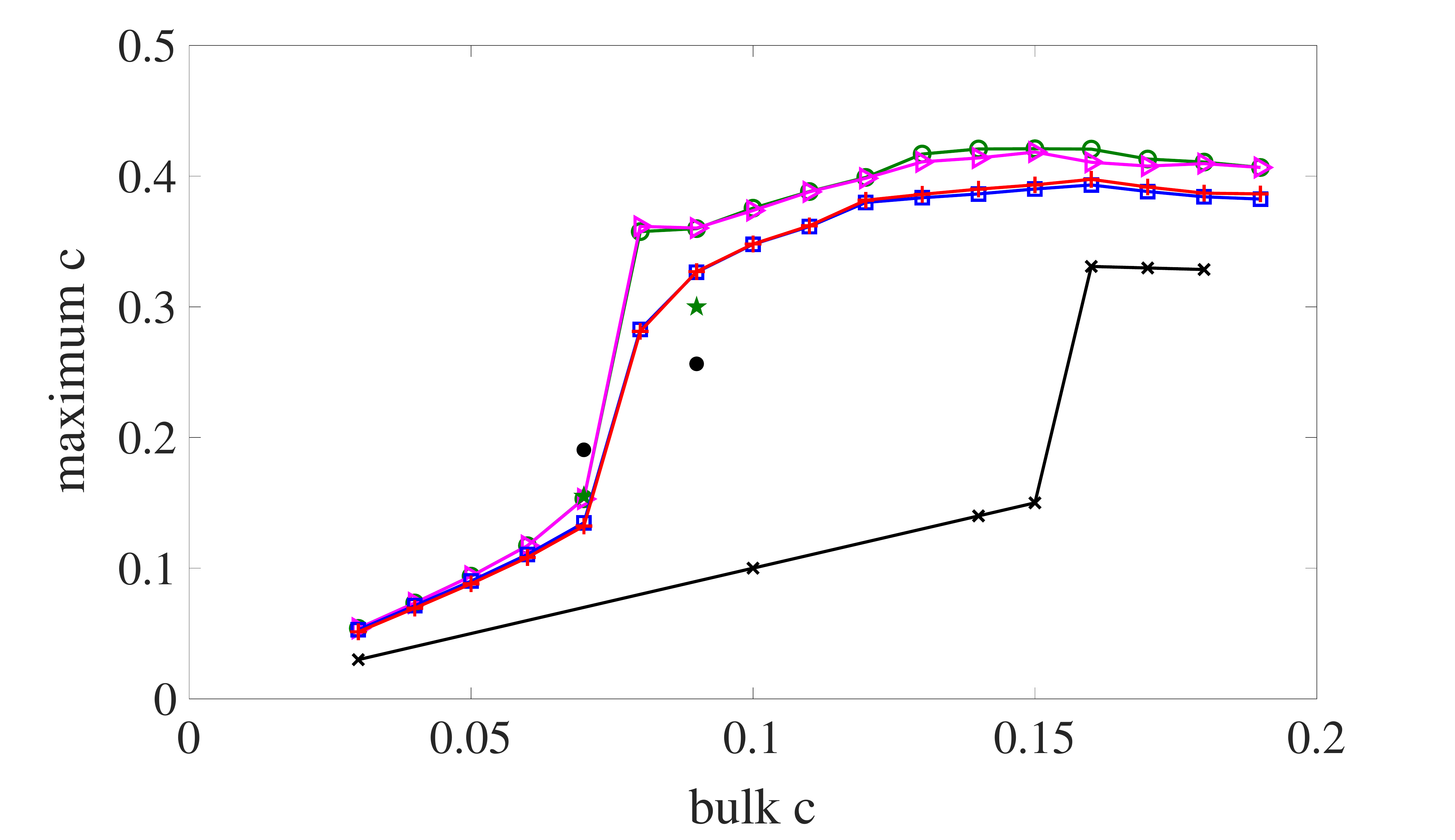}
\caption{Maximum Mn concentration in a system with an edge dislocation for different initially uniform 
bulk Mn concentrations predicted by MPFCM. 
Black crosses:~
\(\nu_{\mathrm{M}}=0\), \(\ssans{C}(0;723)\), \(\kappa_0\), \(M(0.25;723)\).
Red crosses:~\(\nu_{\mathrm{M}}(723)\), \(\ssans{C}(0;723)\), 
\(\kappa_0\), \(M(0.25;723)\). 
Blue squares:~\(\nu_{\mathrm{M}}(723)\), \(\ssans{C}(c;723)\), 
\(\kappa_0\), \(M(0.25;723)\). 
Green circles:~\(\nu_{\mathrm{M}}(723)\), \(\ssans{C}(c;723)\), 
\(\kappa(c;723)\), \(M(0.25;723)\). 
Violet triangles:~\(\nu_{\mathrm{M}}(723)\), \(\ssans{C}(c;723)\), 
\(\kappa(c;723)\), \(M(c;723)\). 
Green stars:~APT results for maximum \(c\) values along the dislocation line in Fe7Mn 
(Figure \ref{fig_APT_7}(d)) and Fe9Mn (Figure \ref{fig_APT_9}(d)). 
Black circles:~results from MCMD.
}
\label{fig_Bulk_Dislo}
\end{figure}

Besides segregation to defects such as dislocations, spinodal decomposition 
is expected to contribute in general as well to spatial fluctuations in 
solute concentration, both in the bulk and at defects. As discussed above, 
MPFCM modeling of the driving force for Mn flux and segregation is based on 
\eqref{equ:PotCheRedCH} for \(\mu\). This as well as the spinodal region 
\(\partial_{c}^{2}\psi\leqslant 0\) is clearly influenced by the \(c\) 
dependence of \(\bm{F}_{\mathrm{M}}\), \(\ssans{C}\), \(\kappa\), and \(M\). 
To discuss this influence in more detail, consider the simulation results in 
Figure \ref{fig_Bulk_Dislo}. 
As shown by these results, segregation to (or from), and spinodal decomposition 
at, a (straight edge) dislocation is most strongly affected by solute misfit, 
and less so by the \(c\)-dependence of \(\kappa(c;723)\); indeed, that of 
\(\ssans{C}(c;723)\) and \(M(c;723)\) is minimal. 
Except in the case of 
no misfit (and hence no segregation; black crosses), note that maximum Mn 
concentration values lie above the bulk values in Figure \ref{fig_Bulk_Dislo} 
from the start due to segregation. For the cases with misfit and \(\kappa\) 
independent of \(c\) (red crosses, blue squares), note that 
(i) less pre-spinodal segregation occurs, 
(ii) spinodal decomposition begins in the core region at slightly lower 
\(c\) than the homogeneous chemical value \(c=0.157\) in 
Figure \ref{fig_freeEn} (right). Account for the \(c\) dependence of \(\kappa\) 
results in (i) more segregation (green circles, violet triangles) and 
(ii) a shift of the spinodal region in the core to higher concentrations. 
In this latter case, the MPFCM results agree well with those from APT 
(green stars) for Fe7Mn (Figure \ref{fig_APT_7}(d)), and less so for 
Fe9Mn (Figure \ref{fig_APT_7}(d)). 

\begin{figure}[H]
\centering
\includegraphics[width=0.85\textwidth]{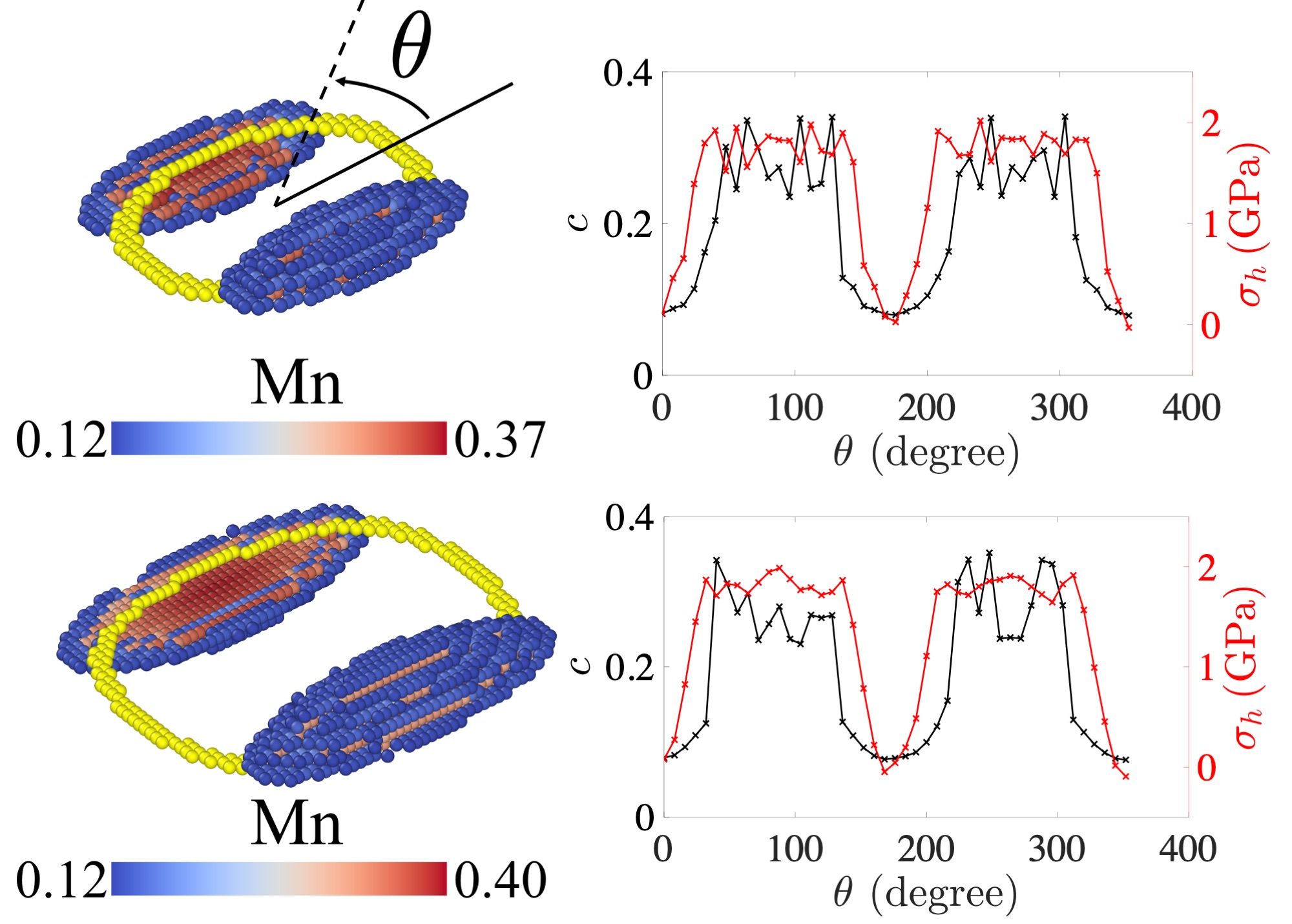}
\caption{MPFCM simulation results for Mn segregation to 
dislocation loops with semi-major axes of 
3.4 nm (top) and 5.0 nm (bottom) (semi-minor axes of 2.4 and 3.5 nm, respectively) 
starting at an initial bulk Mn concentration of 8at.\%. 
Left:~dislocation loop (yellow) and regions containing Mn 
above 12 at.\%. 
Right:~average concentration (black) and hydrostatic stress 
(blue) around the dislocation loop as a function of the angle 
$\theta$. 
}
\label{fig_kink}
\end{figure}

As shown by the results in Figure \ref{fig_PFvsMCMD}, the concentration 
of segregated Mn varies spatially around the dislocation line, related 
in particular to the corresponding variation in the hydrostatic stress 
field. The resulting concentration fluctuations along the (straight edge) line 
are much smaller than those along the line evident in the APT results 
in Figures \ref{fig_APT_7}(d) and \ref{fig_APT_9}(d). 
One possible reason for larger fluctuations could be a change of dislocation 
character along the line (e.g., due to kinks or wavy dislocation structure) and so a change of 
hydrostatic stress along the line. A detailed investigation 
of this possibility is beyond the scope of the current work; as a simple 
"proof of concept", however, consider the case of Mn segregation to 
dislocation loops shown in Figure \ref{fig_kink}. 

As expected, segregation resulting in spinodal decomposition 
is limited to the edge part of the dislocation loop. Note the 
fluctuation of \(c\) of about 10 at.\% in the edge part, i.e., 
even though the hydrostatic stress is relatively uniform. In 
addition, note that the range of concentration fluctuation is 
comparable to that seen in Fe9Mn in Figure \ref{fig_APT_9}(d). 
Based on these results, changes in dislocation character for 
other reasons such as kinking can also be expected to result 
in concentration fluctuations of the type observed in the APT 
results. 


As seen in the modeling results of \cite{ma14071787} and \cite{r2021dislocation}, 
under certain conditions, spinodal decomposition may occur along a 
straight dislocation line. Solute segregation to lines results in an 
initially uniform solute concentration due to 
the uniform stress field along the line. Once this uniform concentration 
reaches the lower spinodal concentration, decomposition occurs between 
the bulk and dislocation line (Fig.~\ref{fig_Bulk_Dislo}). As shown in 
\cite{ma14071787}, if there is insufficient solute in the system for the 
solute concentration to attain the upper spinodal 
concentration during decomposition, however, the uniform 
distribution along the line becomes unstable, and further decomposition occurs. The larger the upper spinodal concentration, and 
the smaller the system size (i.e., solute supply), the more likely this 
is to occur. Note that these results 
hold for the case of uniform, constant solute mobility in the system. 
As shown in \cite{r2021dislocation}, a non-uniform solute mobility, i.e., 
relatively low solute mobility in the bulk, and relative high solute 
mobility along the dislocation line ("pipe diffusion"), can also lead 
to spinodal decomposition along the line. 

For the Fe-Mn system at 723\,K studied in this work, the upper spinodal 
concentration of $c=0.249$ (Fig.~ \ref{fig_freeEn}) is relatively small, 
and the system size is sufficiently large, to result in spinodal decomposition 
of the uniform solute distribution along the line. 
Thus, the insufficient supply of solutes and the resulting 
spinodal breakup along the line is most likely not be the reason for the observed 
concentration fluctuations in the APT experiments. Furthermore, as measured 
in APT, the regions neighboring the dislocation lines do not show significant 
solute depletion, confirming that there is no lack of solute supply. 
Therefore, the system is not kinematically limited and solute 
mobility should not play a role in the spinodal morphology. Given the arguments 
above, we propose that the solute segregation fluctuations around the dislocation 
line observed in the APT for Fe-Mn system are due to structural fluctuations 
and dislocation character changes, rather than the spinodal breakup due to 
insufficient solute \cite{ma14071787} or enhanced mobility around the dislocation 
line \cite{r2021dislocation}. 

\section{Summary and discussion}
\label{sec:DisSum}

Segregation- and spinodal-based Mn enrichment of bulk dislocations in 
BCC Fe has been investigated in the current work with the help of 
comparative modeling and material characterization. On the modeling side, 
both microscopic phase-field chemomechanics (MPFCM) \cite{Svendsen2018} 
and (hybrid) Monte Carlo molecular dynamics (MCMD) \cite{Sadigh2012} have 
been employed. In the former case, the model is calibrated using both the 
MEAM potential (resulting in APFCM) employed in MCMD as well as and using 
Thermo-Calc / CALPHAD data for Fe-Mn. 
Predictions from these models are compared with characterization 
results from atom probe tomography (APT) for two Fe-Mn alloys. 
Both straight edge and screw dislocations, as well as dislocation loops, 
are considered in the simulations. 

In contrast to the case in Volterra (linear elastic) dislocation theory, 
non-linear effects accounted for in both MPFCM and MCMD result in a 
non-zero hydrostatic stress field in screw cores. Being of 
much smaller magnitude than the hydrostatic stress in straight edge cores, 
much less solute segregates to screw than to edge cores, and the segregated 
amount in screw cores is always below spinodal. Additional effects 
captured by MCMD (not accounted for in MPFCM), such as (screw) core spreading, 
or core Kanzaki forces \cite{GLerma2018}, may also play a role in determining 
the (screw) core (hydrostatic) stress state as well. 

Results from MPFCM-based modeling of Mn segregation to, and spinodal 
decomposition at, a straight edge dislocation imply that the concentration 
dependence of the solute misfit distortion \(\bm{H}_{\mathrm{M}}(c;\theta)\) 
and resulting non-linear dependence of the elastic energy density 
\(
\psi_{\mathrm{el}}(c,\bm{F};\theta,\phi)
\) 
on \(c\) have the strongest effect. Segregation and spinodal decomposition 
are less strongly affected by the \(c\)-dependence of \(\kappa(c;723)\) in the 
chemical gradient energy density \(\psi_{\mathrm{gr}}(c,\nabla c;\theta)\). 
In comparison to these two, the influence of the concentration dependence of 
the elastic stiffness \(\ssans{C}(c;723)\) and solute mobility \(M(c;723)\) 
on this behavior is minimal, in the current case of Fe-Mn. 

With respect the maximum amount of Mn segregating to straight edge dislocations, 
the MPFCM results show good agreement with corresponding APT results. On the 
other hand, the current MPFCM model for Fe-Mn predicts little or no fluctuation 
in Mn concentration along the line, in contrast to the APT results. This could 
be related to the fact that MPFCM predicts a uniform hydrostatic stress along 
the line. In this case, segregated Mn is expected to exceed the spinodal value 
in the core uniformly along the line, resulting in the formation of a single 
precipitate, and minimal concentration fluctuations, along the (straight edge) 
line. As shown for example by the dislocation loop case in the current work, 
change in dislocation character (i.e., edge, screw, mixed) along the line, 
however, does result in significant corresponding fluctuations in solute 
concentration. Further investigation of this and the related case of kinked 
dislocations more relevant to the BCC case represent work in progress to 
be reported on in the future. 

\section*{Acknowledgements}

Financial support of the modeling and simulation work reported on here 
in Subproject M5 (M8) of the Priority Program 1713 
"Strong Coupling of Thermochemical and Thermomechanical States in Applied Materials"
of the Deutsche Forschungsgemeinschaft (DFG) is gratefully acknowledged. 
Part of simulations were performed with computing resources granted by RWTH Aachen University under project rwth0482.
PS is grateful to the EPSRC for financial support through the associated programme grant LightFORM (EP/R001715/1).


\end{document}